\documentclass{article}
\usepackage{graphicx} 
\usepackage{amsmath}
\usepackage{enumitem}
\PassOptionsToPackage{unicode}{hyperref}
\PassOptionsToPackage{hyphens}{url}

\usepackage{amsmath,amssymb}
\usepackage{lmodern}
\usepackage{iftex}
\usepackage{authblk}
\ifPDFTeX
  \usepackage[T1]{fontenc}
  \usepackage[utf8]{inputenc}
  \usepackage{textcomp} 
\else 
  \usepackage{unicode-math}
  \defaultfontfeatures{Scale=MatchLowercase}
  \defaultfontfeatures[\rmfamily]{Ligatures=TeX,Scale=1}
\fi
\IfFileExists{upquote.sty}{\usepackage{upquote}}{}
\IfFileExists{microtype.sty}{
  \usepackage[]{microtype}
  \UseMicrotypeSet[protrusion]{basicmath} 
}{}
\makeatletter
\@ifundefined{KOMAClassName}{
  \IfFileExists{parskip.sty}{%
    \usepackage{parskip}
  }{
    \setlength{\parindent}{0pt}
    \setlength{\parskip}{6pt plus 2pt minus 1pt}}
}{
  \KOMAoptions{parskip=half}}
\makeatother
\usepackage{xcolor}
\usepackage{longtable,booktabs,array}
\usepackage{calc} 
\usepackage{etoolbox}
\makeatletter
\patchcmd\longtable{\par}{\if@noskipsec\mbox{}\fi\par}{}{}
\makeatother
\IfFileExists{footnotehyper.sty}{\usepackage{footnotehyper}}{\usepackage{footnote}}
\makesavenoteenv{longtable}
\usepackage{graphicx}
\makeatletter
\def\maxwidth{\ifdim\Gin@nat@width>\linewidth\linewidth\else\Gin@nat@width\fi}
\def\maxheight{\ifdim\Gin@nat@height>\textheight\textheight\else\Gin@nat@height\fi}
\makeatother
\setkeys{Gin}{width=\maxwidth,height=\maxheight,keepaspectratio}
\makeatletter
\def\fps@figure{htbp}
\makeatother
\usepackage[normalem]{ulem}
\ifLuaTeX
  \usepackage{selnolig}  
\fi
\IfFileExists{bookmark.sty}{\usepackage{bookmark}}{\usepackage{hyperref}}
\IfFileExists{xurl.sty}{\usepackage{xurl}}{} 
\hypersetup{
  hidelinks,
  pdfcreator={LaTeX via pandoc}}

\author[1]{Jungin E. Kim}
\author[1]{Yan Wang}
\affil[1]{George W. Woodruff School of Mechanical Engineering, Georgia Institute of Technology, Atlanta, GA 30332 USA}
\title{Quantum Approximate Bayesian Optimization Algorithms
with Two Mixers and Uncertainty Quantification}
\date{October 23, 2023}

\begin{document}

\maketitle

\section*{Abstract}
The searching efficiency of the quantum approximate optimization algorithm is dependent on both the classical and quantum sides of the algorithm. Recently a quantum approximate Bayesian optimization algorithm (QABOA) that includes two mixers was developed, where surrogate-based Bayesian optimization is applied to improve the sampling efficiency of the classical optimizer. A continuous-time quantum walk mixer is used to enhance exploration, and the generalized Grover mixer is also applied to improve exploitation. In this paper, an extension of QABOA is proposed to further improve its searching efficiency. The searching efficiency is enhanced through two aspects. First, two mixers, including one for exploration and the other for exploitation, are applied in an alternating fashion. Second, uncertainty of the quantum circuit is quantified with a new quantum Matérn kernel based on the kurtosis of the basis state distribution, which increases the chance of obtaining the optimum. The proposed new two-mixer QABOA’s with and without uncertainty quantification are compared with three single-mixer QABOA’s on five discrete and four mixed-integer problems. The results show that the proposed two-mixer QABOA with uncertainty quantification has the best performance in efficiency and consistency for five out of the nine tested problems. The results also show that QABOA with the generalized Grover mixer performs the best among the single-mixer algorithms, thereby demonstrating the benefit of exploitation and the importance of dynamic exploration-exploitation balance in improving searching efficiency.

\section{1. Introduction}
Quantum computers are promising in solving large-scale engineering problems such as optimization [\ref{wang_QCOverview}]-[\ref{wang_QuantumWalk}] and simulation [\ref{wang_StochasticDiff}]-[\ref{wang_StochasticDynamics}]. However, the capability of current quantum computers to solve large-scale problems is limited. One reason is that the number of qubits on current quantum computers is small, whereas many qubits are required to represent the large searching space as the number of variables increases. Another reason is that decoherence can occur easily on quantum computers due to environmental noise, which makes useable quantum computing time very short.

One optimization algorithm that was recently developed for noisy intermediate-scale quantum computers is the quantum approximate optimization algorithm (QAOA) [\ref{farhi_QAOA}]. The goal of QAOA is to find optimal rotation angles for a quantum circuit to increase the amplitude of the optimal basis state. The circuit consists of phase-separating and mixer Hamiltonian operators in an alternating fashion. The phase-separating Hamiltonian operator encodes the objective quantity being optimized, whereas the mixer Hamiltonian operator alters the basis state amplitudes. By incorporating a classical optimization algorithm to optimize the rotation angles, this hybrid quantum-classical algorithm can alleviate the decoherence issue.

QAOA is a heuristic optimization algorithm. Its searching efficiency is dependent on both the classical and quantum sides of the algorithm. Several approaches have been proposed to improve the efficiency of QAOA. One approach is to customize the quantum circuit based on the problem. By taking advantage of functional relationships between the objective quantities and the quantum circuit parameters  [\ref{zhou_QAOA}]-[\ref{wang_QAOAmaxcut}], or exploiting the symmetry of the objective function [\ref{shaydulin_QAOA}], the size of the rotation angle searching space is reduced. Another approach is to initialize the quantum circuit other than the uniform superposition. Examples of initialization strategies are warm-start preprocessing [\ref{tate_warmQAOA}] and Dicke states [\ref{cook_QAOAz}], which are based on the prior knowledge about the potential solutions. A third approach is to define the phase-separating Hamiltonian with an Ising model of higher order than the quadratic form [\ref{campbell_QAOA}]. This binary native encoding scheme can achieve higher quality solutions than the unary and binary reduced encoding schemes.

The most critical component of QAOA is the mixer
operator, which determines how the state of the quantum
system evolves. Several mixers have been proposed to
improve the searching efficiency of QAOA. Some mixers
such as the XY mixer [\ref{hadfield_QAOAz}]-[\ref{wang_XY}] and the continuous-time
quantum walk mixer [\ref{marsh_quantum}] restrict the search space to the states which satisfy problem constraints. The generalized Grover mixer [\ref{bartschi_grover}] increases the amplitudes of basis states associated with improved objective values. The continuous-time quantum walk and generalized Grover mixers are also
combined to improve the exploration-exploitation balance
[\ref{ywang_QABOA}]. A free-axis mixer [\ref{govia_freedom}] and an entangled gate mixer [\ref{chen_entangle}] induce a wider exploration of the basis states.

Since mixers affect the searching efficiency of QAOA, the design of mixers should incorporate the exploration-exploitation balance. Exploration involves finding the global optimal solutions in new regions of the search space, whereas exploitation involves further improving the current best solution locally. Maintaining a good balance between exploration and exploitation is important for high searching efficiency. Over-exploration may miss the opportunity to further improve the solution, whereas over-exploitation has the risk of being trapped at a local optimum.

One way to achieve the exploration-exploitation balance is to incorporate two mixers in the quantum circuit. The first mixer emphasizes exploration by altering the basis state amplitudes. The second mixer performs exploitation by increasing the amplitudes of better solutions. In the recent quantum approximate Bayesian optimization algorithm (QABOA) [\ref{ywang_QABOA}], the continuous-time quantum walk mixer enhances exploration, whereas the generalized Grover mixer improves exploitation. This two-mixer design can effectively improve the balance. In addition, Bayesian optimization (BO) is utilized in QABOA for the classical optimization side. BO is a surrogate-based global optimization approach that can improve the sampling efficiency of optimization. Gaussian process regression (GPR) model is the most commonly used surrogate for sequential sampling in BO. An acquisition function is constructed based on the surrogate to guide the search into the most promising regions. The acquisition function can also be designed to achieve a good exploration-exploitation balance.

The above research efforts are to improve the searching efficiency of the QAOA. The uncertainty associated with quantum circuit evaluation and quantum noise, however, was not considered. The probabilistic nature of measurements leads to non-deterministic and random results. The quantum noise is due to the errors and decoherence of the computer system. Incorporating uncertainty is important to improve consistency and speed of convergence.

In this work, an extension of QABOA is proposed to further improve its performance. The searching efficiency is improved through two aspects. First, two mixers, including one for exploration and the other for exploitation, are applied in an alternating fashion. Pauli-X, XY, and quantum walk are examples of exploration mixers. The generalized Grover mixer is for exploitation through amplitude amplification. The new algorithm with two mixers is referred to as TM-QABOA. TM-QABOA can be used to solve both discrete and mixed-integer optimization problems, where BO is performed to optimize both rotation angles and continuous variables in the original objective function. Second, uncertainty of the quantum circuit is incorporated in the searching. The uncertainty is quantified by introducing a new quantum Matérn kernel based on the estimated kurtosis, or peak sharpness, of the basis state distribution. The version of TM-QABOA which incorporates the uncertainty is referred to as uTM-QABOA. The results of uTM-QABOA show that the searching efficiency can be improved with the consideration of uncertainty. The number of quantum circuit runs in the proposed QABOA’s is reduced in two aspects. First, an improved exploration-exploitation balance with the two mixers can help obtain the optimal solution faster. Second, surrogate-based BO can reduce the number of objective function evaluations which correspond to the quantum circuit runs.

The remainder of the paper is structured as follows. In
Section 2, the relevant work of improving QAOA efficiency
is reviewed. The sources of uncertainty in quantum
optimization algorithms and the methods to reduce quantum
noise are introduced. The proposed TM-QABOA
methodology is described in Section 3. In Section 4, the
TM-QABOA is demonstrated and evaluated with nine
optimization problems. The problems include a MaxCut problem, three weighted MaxCut problems, lattice protein folding, HeH$^+$ potential
energy minimization, as well as the designs of a welded
beam, a speed reducer, and a pressure vessel. The first five are discrete optimization problems, whereas the last four are mixed-integer problems. Future extensions of TM-QABOA are discussed in Section 5.

\section{2. Relevant Work}
\subsection{2.1. Existing Mixers to Improve the QAOA Efficiency}
The goal of QAOA is to optimize rotation angles for the
quantum circuit to increase the optimal basis state amplitude. Given an $n$-qubit quantum system, the QAOA quantum circuit with a depth of $p$ is defined as

\begin{equation}\label{QAOA_QC}
    |\psi\rangle = U_B(B, \beta_p)U_C(C, \gamma_p){\cdots}U_B(B, \beta_1)U_C(C, \gamma_1)|+\rangle^{{\otimes}n}
\end{equation} 

\noindent where $|+\rangle^{{\otimes}n}$ is the initial state of the $n$-qubit system with uniform superposition. $|\psi\rangle$ is the final state. The operator $U_C(C, \gamma)$ is the phase-separating Hamiltonian operator with phase-separating Hamiltonian $C$ and rotation angle $\gamma$. The mixer operator $U_B(B, \beta)$ is defined with mixer Hamiltonian $B$ and rotation angle $\beta$. The quantum circuit alternates between $U_C$ and $U_B$ for $p$ repetitions. The operator $U_C$ encodes the objective quantity being optimized, whereas $U_B$ perturbs the quantum system to change the system's state.

The choice of $B$ affects how the quantum system evolves.
Various mixers have been proposed to
improve the QAOA searching efficiency. One type of mixer is
the XY mixer [\ref{wang_XY}], which explores basis states that are
topologically connected as graphs. With this graph structure,
the XY mixer ensures that only feasible states are explored.
XY mixers are used in the quantum alternating operator ansatz
[\ref{hadfield_QAOAz}], which results in a faster search process than the original
QAOA with Pauli-X mixers. The two mixers have been
compared in solving different problems, such as k-vertex
cover [\ref{cook_QAOAz}], portfolio optimization [\ref{brandhofer_QAOA}], and extractive text
summarization [\ref{niroula_constrained}]. Furthermore, the phase-separating and
XY mixer operators are combined into a two-parameter
unitary operator [\ref{larose_mixerphaser}]. As a result, the quantum circuit depth is
reduced. Another mixer is the continuous-time quantum walk
[\ref{marsh_quantum}]. This mixer is similar to the XY mixer by which it represents the feasible basis state space in combinatorial problems as graphs. With this mixer, the quantum system’s
evolution follows the problem constraints.

The generalized Grover mixer [\ref{bartschi_grover}]-[\ref{ywang_QABOA}] is based on Grover’s
algorithm [\ref{grover_database}] originally developed for the unstructured
database search problem. In the generalized Grover operator,
a rotation parameter is introduced which replaces the fixed $\pi$
rotation angle. The generalized Grover operator can be used as
a mixer in QAOA. Recently, Wang [\ref{ywang_QABOA}] proposed QABOA. The quantum circuit consists of the generalized
Grover operator and the continuous-time quantum walk as two
mixers which are applied together. BO is applied to
optimize the rotation angles. The QABOA quantum circuit is designed to improve the exploration-exploitation balance.

Other QAOA mixers have been devised. Govia et al. [\ref{govia_freedom}]
proposed a free-axis mixer which performs qubit rotations
about any axis in the XY plane. The mixer results in higher
approximation ratios than the original QAOA. Chen et al. [\ref{chen_entangle}]
incorporated quantum entanglement of Pauli-X and Pauli-Y
mixers to improve convergence. Similarly, Yu et al. [\ref{yu_QAOAbias}]
proposed the adaptive bias mixer which combines the Pauli-X
and parameterized Pauli-Z gates to reduce the circuit depth.
Chancellor [\ref{chancellor_domain}] devised a mixer which implements coupling
between qubits so that the total number of qubits can be
reduced.

Currently, most of the QAOA mixers either improve the
searching efficiency by shrinking the search space with
constraints, or improve the convergence towards the global
optimum by exploring a larger search space. However,
research efforts to improve the exploration-exploitation
balance are limited. A dynamic exploration-exploitation
balance is important to improve the searching efficiency. An
algorithm which over-explores can miss the global optimum,
whereas an algorithm which over-exploits can remain trapped
at a local optimum. Here, an extension of QABOA is proposed
to improve the exploration-exploitation balance with an
alternating sequence of two mixers.

\subsection{2.2. Sources of Uncertainties in Quantum Optimization}
\noindent One source of the uncertainty in quantum optimization is
quantum noise or computer error, which causes the state of a quantum system to
change randomly. In variational algorithms such as
QAOA, quantum noise results in the rotation angles deviating
from optimal values [\ref{xue_QAOAnoise}]-[\ref{marshall_QAOAnoise}]. A common approach to reduce
quantum noise is quantum error correction [\ref{roffe_errorcorrect}], where the
state of the quantum system is encoded with a redundant
number of qubits. Examples of quantum error correction
algorithms include Shor’s method [\ref{shor_dechoerence}], stabilizer codes [\ref{gottesman_stabilizer}],
surface codes [\ref{kitaev_anyon}]-[\ref{dennis_topomemory}], and bosonic codes [\ref{cai_bosonic}]. Another
approach is to implement a noise model which captures noise.
For example, an extension of the variational quantum
eigensolver was proposed [\ref{muller_VQE}], where a GPR model is applied to reduce noises in objective values. With
the GPR, the number of iterations to find near-optimal
solutions decreases.
\pagebreak

Another source is the non-deterministic nature of basis state measurements. Given the same rotation angles in the QAOA circuit, obtaining the same optimal solution from multiple runs is not guaranteed. To improve the consistency and speed of convergence, the random error associated with the amplitude estimation for the optimal solution needs to be estimated.

In general, when measurements are taken, the uncertainty associated with the amplitudes is the combined effect from both computer error and random error. The quantification of the uncertainty associated with the amplitudes is currently unexplored in QAOA’s. In this work, a method of quantifying the uncertainty is proposed.

\section{3. Proposed Quantum Approximate Bayesian Optimization Algorithms with Two Mixers (TM-QABOA)}

\noindent In this section, the proposed TM-QABOA methodology is
described. In TM-QABOA, BO is combined with a quantum
circuit consisting of three types of unitary operators arranged
in an alternating fashion. FIGURE 1 illustrates the TM-QABOA architecture, where BO is the outer optimization loop,
and the quantum circuit is the inner one. The
three types of unitary operators include phase-separating
Hamiltonian operators $U_C$’s, mixer Hamiltonian operators
$U_B$’s, and generalized Grover mixers $U_G$’s. First, each $U_C$ is
associated with a rotation angle $\gamma_i$
($i$ = 1, ..., $p$) and phase-separating Hamiltonian $C(\textbf{x}_c)$, where $C(\textbf{x}_c)$
depends on a collection of continuous variables $\textbf{x}_c$. Second, each $U_B$ is associated with a
rotation angle $\beta_i$
($i$ = 1, ..., $p$) and a mixer Hamiltonian $B$.
Lastly, each $U_G$ is associated with a rotation angle $\theta_i$
($i$ =
1, ..., $p$). BO is used to optimize the rotation angles. TM-QABOA can also be applied to mixer-integer optimization
problems, where $\textbf{x}_c$
is optimized with BO. The GPR surrogate models a Gaussian distribution of possible objective values at a sample point. GPR is defined with a covariance kernel function which quantifies the similarity between solutions.

\bigskip
\begin{center} 
\includegraphics[width=4.5in,height=2.4in]{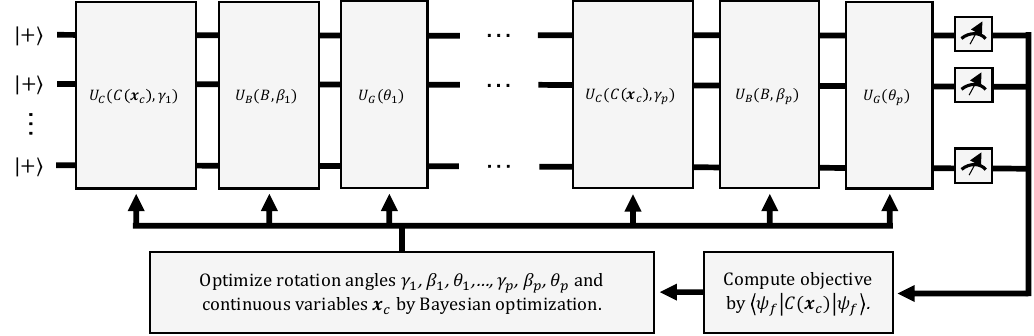}
\textbf{FIGURE 1.} Proposed TM-QABOA architecture.
\end{center}
\bigskip
\pagebreak

In addition, to quantify the uncertainty associated with the amplitudes of optimal solutions, a second version of TM-QABOA is proposed, referred to as uTM-QABOA. In the second version, the GPR is defined with a novel kernel function called the quantum Matérn (QM) kernel, which is the sum of the Matérn kernel and a new quantum kurtosis kernel. The quantum kurtosis kernel is inversely related to the kurtosis, or peak sharpness, of the estimated basis state distribution.

The remainder of this section is structured as follows. The
quantum circuit architecture is described in Section 3.1. The BO
process is described in Section 3.2. The QM kernel is
formulated in Section 3.3. The procedure of uTM-QABOA is
summarized in Section 3.4.

\subsection{3.1. Quantum Circuit Architecture} \label{QC}
\noindent The TM-QABOA quantum circuit with depth $p$ is defined as 

\begin{equation} \label{TMQABOA_QC}
\begin{gathered}
    |\psi\rangle = U_G(\theta_p)U_B(B, \beta_p)U_C(C(\textbf{x}_c), \gamma_p)\cdots\\
    U_G(\theta_1)U_B(B, \beta_1)U_C(C(\textbf{x}_c), \gamma_1)|+\rangle^{{\otimes}n}.
\end{gathered}
\end{equation}

\noindent The initial system state $|+\rangle^{{\otimes}n}$ is prepared by initializing all
$n$ qubits as $|{0}\rangle$ and then applying the Hadamard gate to each qubit. This results in a uniform superposition where all basis
states have equal amplitudes of $2^{-n/2}$. The initial
superposition is uniform since the optimal solution is
unknown at first.

The first operator is $U_C$, which encodes the objective quantity being optimized. The operator $U_C$ is defined as

\begin{equation} \label{U_C}
    U_C(C(\textbf{x}_c), \gamma) = e^{-i{\gamma}C(\textbf{x}_c)}.
\end{equation}

\noindent Given an objective function $f(q_1, ..., q_n, \textbf{x}_c)$ where $q_i$ ($i$ = 1, ..., $n$) is a binary variable, $C(\textbf{x}_c)$ is computed by performing a transformation on $f(q_1, ..., q_n, \textbf{x}_c)$. The transformation is defined as [\ref{wang_QAOAmaxcut}]

\begin{equation} \label{pauliZsub}
    q_i \rightarrow \frac{1}{2}(1-Z_i)
\end{equation}

\noindent where $Z_i$ is the Pauli-Z gate acting on the $i^{th}$ qubit.

The second operator $U_B$, which is the first mixer in the quantum circuit, perturbs the system from one state to another. The operator $U_B$ is defined as

\begin{equation} \label{U_B}
    U_B(B, \beta) = e^{-i{\beta}B}
\end{equation}

\pagebreak
\noindent where

\begin{equation} \label{pauliXmixer}
    B = \sum_{i=1}^{n} X_i
\end{equation}

\noindent is a collection of Pauli gates acting on all qubits. The purpose of $U_B$ is to help explore the basis state space.

The third operator $U_G$, which is the second mixer in the quantum circuit, performs amplitude amplification on basis states with improved objective values. The operator $U_G$ is a combination of two reflection operators, defined as

\begin{equation} \label{U_G}
    U_G(\theta) = {U_R}U_S(\theta)
\end{equation}

\noindent where

\begin{equation} \label{U_S}
    U_S(\theta) = I - (1 - e^{i\theta})\sum_{u \in S}|u\rangle{\langle}u|
\end{equation}

\noindent shifts the phases of the target basis states by a rotation angle $\theta$. Here, $I$ is the identity matrix and $S$ is the set of solutions which have better objective values than the current best objective value $f^{*}$ found so far. Next,

\begin{equation} \label{U_R}
    U_R = U_H(I - 2|\textbf{0}\rangle{\langle}\textbf{0}|)U_H
\end{equation}

\noindent reflects all basis states about the average amplitude. Here, $|\textbf{0}\rangle$
is the basis state with all qubits set to $|0\rangle$ and $U_H$ is the tensor
product of $n$ Hadamard gates.

To implement the generalized Grover mixer, two quantum registers can be utilized, similar to [\ref{gilliam_GAS}]. The first register encodes the solution of the optimization problem, while the second register is used to store the difference between the current objective value and the best solution found so far. All qubits in the second register are initialized as $|0\rangle$. The generalized Grover mixer consists of the following operations. First, Hadamard gates are applied on the second register to create a superposition of basis states. Next, several controlled gates shift the phase of each basis state by an angle proportional to the difference between the current and best objective values. The phase shifts are controlled by the first register. Third, the inverse quantum Fourier transform is applied on the second register, which converts the phase encoding of the difference between objective values into a binary representation. As part of this representation, one of the qubits in the second register indicates the sign of the difference between objective values, which helps determine the solution set $S$ in Eq. (\ref{U_S}). Next, the phases of target basis states are shifted by an angle $\theta$. This operation is controlled by the qubit representing the sign of the value difference. After the phases of the target basis states are shifted with $U_S$, $U_R$ is performed on the first register. The second register is then uncomputed to reset the qubits to $|0\rangle$ in order to release the register entanglement.

After repeating the sequence of $U_C$, $U_B$, and $U_G$ for $p$ times,
the measurement operators collapse $|\psi\rangle$ into a single basis
state $|\psi_f\rangle$, which is the optimal solution that the algorithm
identifies. The objective value corresponding to $|\psi_f\rangle$ is
computed from the expectation of $C(\textbf{x}_c)$, defined as

\begin{equation} \label{Hamil_expect}
    f(\textbf{x}_c, |\psi_f\rangle) \approx {\langle}\psi_f|C(\textbf{x}_c)|\psi_f\rangle.
\end{equation}

The TM-QABOA quantum circuit is designed with two
mixers to improve the searching efficiency by inducing a
dynamic exploration-exploitation balance. The first mixer $U_B$
alters the basis state amplitudes for exploration of different
basis states, whereas the second mixer $U_G$ helps exploit basis states
with improved objective values. Alternating between both
mixers can potentially increase convergence towards
optimality by which basis states with improved objective
values are more likely to be measured.

\subsection{3.2. Bayesian Optimization}
\noindent In the proposed TM-QABOA framework, surrogate-based BO is used to
optimize the quantum circuit rotation angles $\gamma_i$, $\beta_i$, $\theta_i$
($i$ = 1, ..., $p$), and the
continuous variables $\textbf{x}_c$ in the original objective function. In contrast to gradient-based local optimization methods, BO is a surrogate-based global optimization approach. It does not require the objective function to be explicitly known. A surrogate of the objective function is constructed to improve the searching efficiency. BO does not require the derivatives of the objective function to be computed. The number of objective function evaluations can be significantly reduced. Furthermore, uncertainty of the objective is incorporated into the GPR surrogate model so that the robustness of the optimization is improved. Here, the uncertainty associated with quantum computation is quantified based on BO by defining a new QM kernel.

In BO, the GPR surrogate model is
trained and updated on a collection of sample points. The GPR generates a distribution of possible objective functions which fit the known sample points. The mean and standard deviation of the surrogate model are used to determine another sample point to add to the collection. The selected sample points guide the search process towards the optimal solution.

First, an initial collection of sample points are obtained from either a pre-existing dataset or random sampling from the
search space. The GPR is then constructed based on the sample points.
Given a sample point $\textbf{x}$ in the search space, the objective value $f(\textbf{x})$ follows a Gaussian distribution, defined as

\begin{equation} \label{GPR}
    f(\textbf{x}) \sim GP(m(\textbf{x}), K(\textbf{x}, \textbf{x}'))
\end{equation}

\noindent where $m(\textbf{x})$ is the mean function and $K(\textbf{x}, \textbf{x}')$ is a covariance kernel function for two sample points $\textbf{x}$ and $\textbf{x'}$. A common kernel function is the Mat\'ern kernel, defined as

\begin{equation} \label{Matern}
K_M(\textbf{x}, \textbf{x}') = \frac{1}{\Gamma(\nu)2^{\nu - 1}}(\frac{\sqrt{2\nu}}{l}d(\textbf{x}, \textbf{x}'))^{\nu}K_\nu(\frac{\sqrt{2\nu}}{l}d(\textbf{x}, \textbf{x}'))
\end{equation}

\noindent where $\nu$ is the smoothness hyperparameter, $\Gamma$ is the gamma
function, $l$ is the length scale, $d$ is the Euclidean distance
function, and $K_\nu$
is a modified Bessel function of the second kind. The Mat\'ern kernel is appropriate for an objective
function that changes significantly with small variations in the input variables. A small value of $\nu$ prevents over-smoothing of the objective function.

During each BO iteration, a new sample point is selected by
maximizing an acquisition function. The acquisition function
quantifies how promising a sample point is in finding the
optimal solution. One example of an acquisition function is the
upper confidence bound (UCB) function. For minimization, it is defined as

\begin{equation} \label{UCB}
    A_{UCB}(\textbf{x}) = {\alpha}\sigma(\textbf{x}) - \mu(\textbf{x})
\end{equation}

\noindent where $\alpha$ > 0 is a tradeoff parameter, $\sigma(\textbf{x})$ is the standard deviation of the objective value, and $\mu(\textbf{x})$ is the average objective value at $\textbf{x}$. The value of $\alpha$ determines whether
exploration or exploitation is favored. For a small value of $\alpha$,
the search process favors sample points with small objective
values. For a large value of $\alpha$, the search process favors sample
points with large levels of uncertainty. After the next sample point is determined, the objective value is calculated, and the surrogate is updated with the new sample point. The optimization process continues until convergence criteria are met.

The kernel function affects the accuracy and precision of the GPR in estimating possible objective functions. For the TM-QABOA, the objective value is non-deterministic since a basis state is measured at random. The kernel function must incorporate the uncertainty associated with the basis state measurements to improve the precision of the objective value estimation. The novel QM kernel function which incorporates the uncertainty of the basis state measurements is presented in the following sub-section.

\subsection{3.3. Quantum Mat\'ern Kernel Function}
\noindent The proposed QM kernel function is defined as

\begin{equation} \label{K_QM}
    K_{QM} = K_M + K_Q
\end{equation}

\noindent where $K_M$ is the Mat\'ern kernel defined in Eq. (\ref{Matern}), and

\begin{equation} \label{K_Q}
    K_Q(\textbf{x}, \textbf{x}') = 
    \begin{cases}
    {\omega}(\kappa + \varepsilon)^{-2} & \text{($\textbf{x} = \textbf{x'}$)}\\
    0 & \text{($\textbf{x} \neq \textbf{x'}$)}
    \end{cases}
\end{equation}

\noindent is a new quantum kernel as a function of the Pearson kurtosis
$\kappa$ of the estimated basis state distribution, where $\omega$ > 0 is a
scaling hyperparameter, and $\varepsilon$ is a small positive value to
prevent the division by zero. 

Suppose $z$ is an integer-valued random variable such that $z =$ 0, 1, ..., $2^n - 1$ corresponds to the $n$-qubit basis states $|00...00\rangle, |00...01\rangle, ..., |11...11\rangle$, respectively. The Pearson kurtosis for $z$ is calculated as

\begin{equation} \label{kurtosis}
    \kappa = \frac{\xi}{\sigma^4}
\end{equation}

\noindent where $\xi = {\int}(z-\mu)^{4}p(z)dz$ is the fourth central moment, $\mu$
is the mean, and $\sigma$ is the standard deviation. The value of $\kappa$
indicates how sharp the peak of the distribution is. A very large
value of $\kappa$ for the basis state distribution means that a few
amplitudes are significantly larger than the rest. In this case,
the uncertainty associated with the quantum circuit
measurement is small. As $\kappa$ increases, the uncertainty of the
basis state measurement decreases quadratically.

In the QM kernel function, the key aspect for improving the searching efficiency is the addition of $K_Q$. Since the basis state
measurements are stochastic in nature, the optimization
process is susceptible to the variation in the objective
evaluations. For this reason, $K_Q$ depends on the kurtosis which represents the consistency of basis state measurements.
Adding $K_Q$ to $K_M$ allows the objective values at sample points
to vary, which increases the number of possible objective
functions described by the distribution of the probabilistic GPR. In turn, there is a higher chance that the objective values at the new sample points become closer to the optimal
solution. Furthermore, $K_Q$ depends on the scaling
hyperparameter $\omega$, which captures the extent of the variation in objective values at the known sample points. As the search process proceeds, $\omega$ is tuned with the L-BFGS-B algorithm so that the optimal solution can be obtained in fewer iterations. Therefore, the convergence towards the optimal solutions can be accelerated. The proposed QM kernel is used in the surrogate model of the new uTM-QABOA.

\subsection{3.4. Proposed uTM-QABOA Procedure}
\noindent In the proposed uTM-QABOA, Monte Carlo sampling is
performed on the same quantum circuit of TM-QABOA in
FIGURE 1 to estimate the basis state distribution so that the kurtosis can be calculated.

\pagebreak
\begin{center}
\textbf{TABLE 1.} uTM-QABOA procedure.\\
\includegraphics[width=15in,height=7in]{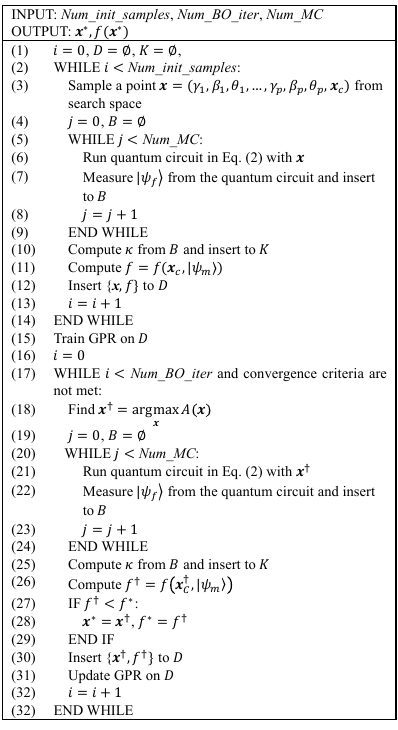}
\end{center}
\pagebreak

TABLE 1 shows the uTM-QABOA procedure. First, initial
samples of the rotation angles $\gamma_i$, $\beta_i$, $\theta_i$
($i$ = 1, ..., $p$) and
continuous variables $\textbf{x}_c$ are randomly acquired from the
search space. For each initial sample, the quantum circuit
defined in Eq. (\ref{TMQABOA_QC}) is performed for multiple runs. During each
run, a basis state $|\psi_f\rangle$ is measured and the result is stored in
the set $B$. The kurtosis $\kappa$ of the basis state distribution is computed after those runs and stored in the set $K$. The
objective value $f$ of each sample point $\textbf{x}$ is computed with the
most frequently measured basis state $|\psi_m\rangle$. The sample $\{\textbf{x}, f\}$
is added to the dataset of known sample points $D$. The GPR
with the QM kernel is then trained on the initial sample points.
The GPR hyperparameters are tuned during the training
process. After the initial GPR model is constructed, the next
sample point $\textbf{x}^\dag$
is determined by maximizing the acquisition
function $A(\textbf{x})$ in each iteration. The quantum circuit is
performed for several runs to approximate the basis state
distribution, from which $\kappa$ is computed. The new objective
value $f^\dag$
is also computed with $|\psi_m\rangle$. The current best sample $\textbf{x}^*$
and its objective value $f^*$
are then updated if the new
objective value is better than the current best one. The new
sample $\{\textbf{x}^\dag, f^\dag\}$ is added to $D$, and the GPR is updated. The
optimization process continues until convergence criteria are met.

Multiple runs of the quantum circuit are needed to compute
the kurtosis of the basis state distribution. Given the
probabilistic nature of the basis state measurements, QAOA’s
are usually run multiple times to receive reliable results.
Therefore, uTM-QABOA does not increase the computational
cost significantly.

\section{4. Optimization Examples}
\noindent In this section, TM-QABOA and uTM-QABOA are
demonstrated with nine optimization problems. The first problem,
discussed in Section 4.1, is unweighted MaxCut with a complete graph of six vertices. The next three problems, discussed in Section 4.2, are weighted MaxCut problems with complete graphs of five vertices. The fifth problem, discussed in Section 4.3, is lattice protein folding. The sixth problem, discussed in Section 4.4, is potential energy minimization of the ionic helium hydride (HeH$^+$) molecule. The seventh problem, discussed in Section 4.5, is the design of a welded beam. The eighth problem, discussed in Section 4.6, is the design of a speed reducer. The ninth problem, discussed in Section 4.7, is the design of a cylindrical pressure vessel. The first five examples involve discrete search spaces, whereas the last four examples involve mixed-integer search spaces.

For all nine examples, TM-QABOA and uTM-QABOA are compared with three single-mixer QABOA’s, which include X-QABOA, XY-QABOA, and GM-QABOA, where the same BO framework is applied to optimize rotation angles and continuous variables. X-QABOA consists of Pauli-X mixers, XY-QABOA consists of XY-graph mixers, and GM-QABOA consists of generalized Grover mixers.

\subsection{4.1. MaxCut Problem with 6-Complete Graph}
\noindent Let $G(V, E)$ be a graph where $V$ and $E$ are the sets of vertices and edges of $G$, respectively. Suppose $V$ is partitioned into two non-overlapping subsets $V_0$ and $V_1$. The objective of the MaxCut problem is to maximize the number of edges in $E$ connecting vertices between $V_0$ and $V_1$. The objective function is defined as 

\begin{equation} \label{maxcut_obj}
    f(\textbf{q}) = \sum_{\{i,j\} \in E} (q_i + q_j - 2{q_i}{q_j})
\end{equation}

where $q_i$ ($i$ = 1, ..., $n$) indicates whether the vertex $i$ belongs to $V_0$ (if $q_i = 0$) or $V_1$ (if $q_i = 1$). $n$ is the total number of vertices. $\textbf{q} = (q_1, q_2, ..., q_n)$ is a combinatorial choice of vertices and $(i, j)$ denotes the edge between vertices $i$ and $j$. Since there are $n$ binary variables, $n$ qubits are needed to solve the problem.

In this example, the MaxCut problem is solved with a complete graph of six vertices. Each of the five algorithms is performed twenty times to determine the average convergence performances. All algorithms consist of six rotation angles to keep the dimension of the search space the same.  For X-QABOA, XY-QABOA, and GM-QABOA, two rotation angles $\gamma_i$ and $\beta_i$ ($i = 1, 2, 3$) are repeated three times. For TM-QABOA and uTM-QABOA, three rotation angles $\gamma_i$, $\beta_i$, and $\theta_i$ ($i = 1, 2$) are repeated twice. All rotation angles range from 0 to 2$\pi$.

The initial dataset to construct the initial GPR model consists of three sample points. Each sample point corresponds to six values of the rotation angles. The same initial dataset is used to initialize the GPR models in all five algorithms for the purpose of comparison. The initial values of $\gamma_1$, $\beta_1$, $\gamma_2$, and $\beta_2$ are the same for all algorithms. The initial values of $\theta_1$ and $\theta_2$ are the same for TM-QABOA and uTM-QABOA. For X-QABOA, XY-QABOA, and GM-QABOA, $\gamma_3$ and $\beta_3$ are initialized as zeroes to maintain the same circuit depth. Each run occurs for 10 BO iterations since the global optimum can be quickly obtained from 64 possible solutions. The GPR is defined with the Matérn kernel for all algorithms except uTM-QABOA. The value of $\nu$ is set to 0.5 for both Matérn and QM kernels to model large objective value changes with small rotation angle changes. The acquisition function for all algorithms is UCB, which allows for the ease of controlling the balance between exploration and exploitation by setting the $\alpha$ value in Eq. (\ref{UCB}). Here, $\alpha$ is set to 1. The acquisition function is maximized by simulated annealing for 50 iterations. 

FIGURE 2(a) and (b) show the averages and standard deviations of the best observed objective values, respectively. In FIGURE 2(a), GM-QABOA is the fastest to converge towards the global maximum value of 9. By the first iteration, all twenty runs of GM-QABOA result in the global optimal solution. Meanwhile, uTM-QABOA converges the slowest towards the global optimum.  However, at the early stage of optimization, it is possible for uTM-QABOA to converge towards the global optimum. For instance, after the first iteration, uTM-QABOA results in an average best value of 7.4 with a standard deviation of 1.96. The global optimum is within one standard deviation of the average value. The performance of TM-QABOA is similar to those of other algorithms.

\bigskip
\begin{center}
\includegraphics[width=6.5in,height=2.4in]{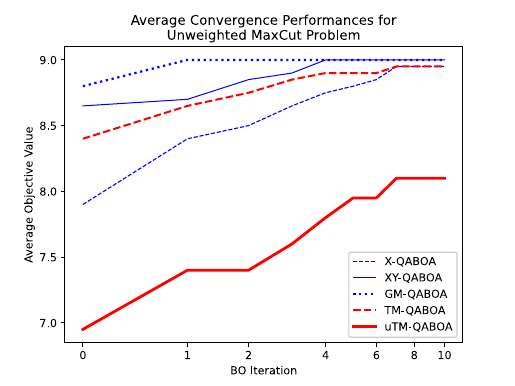}\\
(a)\\
\smallskip
\includegraphics[width=6.5in,height=2.4in]{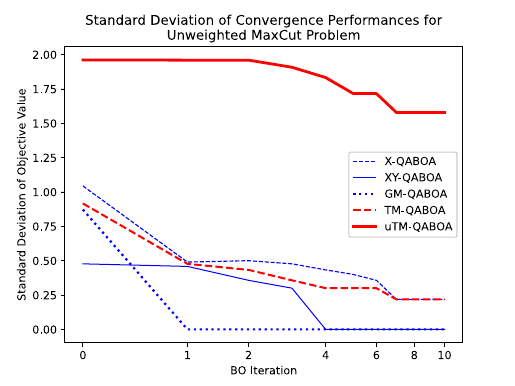}\\
(b)\\
\smallskip
\textbf{FIGURE 2.} (a) Averages and (b) standard deviations for the best observed objective values for the MaxCut problem.
\end{center}
\bigskip

\subsection{4.2. Weighted MaxCut Problem with 5-Complete Graphs}
\noindent A generalized version of the MaxCut problem is the weighted MaxCut problem, where each edge ($i$, $j$) has a weight value $W_{ij}$. The objective of the weighted MaxCut problem is to maximize the total sum of weights associated with edges connecting vertices between $V_0$ and $V_1$. The objective function is defined as

\begin{equation} \label{weightedmaxcut_obj}
    f(\textbf{q}) = \sum_{\{i,j\} \in E} W_{ij}(q_i + q_j - 2{q_i}{q_j}).
\end{equation}

This function is similar to Eq. (\ref{maxcut_obj}), except that each term in the sum is multiplied by $W_{ij}$.

In this example, the weighted MaxCut problem is solved on three instances of a complete graph with 5 vertices. The graph is illustrated in FIGURE 3. Three sets of weight values are randomly generated as listed in TABLE 2. The global maximum values are 32.3 for Graph 1, 36.4 for Graph 2, and 38.5 for Graph 3. Since the problem involves five binary variables $q_i$ ($i = 1, 2, 3, 4, 5$), five qubits are needed to solve the problem.

\begin{center}
\includegraphics[width=4in,height=2.4in]{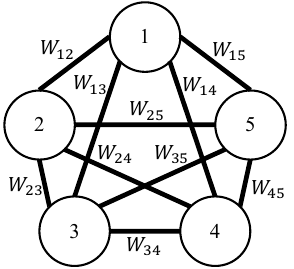}\\
\textbf{FIGURE 3.} 5-complete graph with edge weights.
\end{center}
\bigskip

\begin{center}
\textbf{TABLE 2.} Randomly generated edge weight values for three 5-complete graphs.\\
\includegraphics[width=4in,height=2.4in]{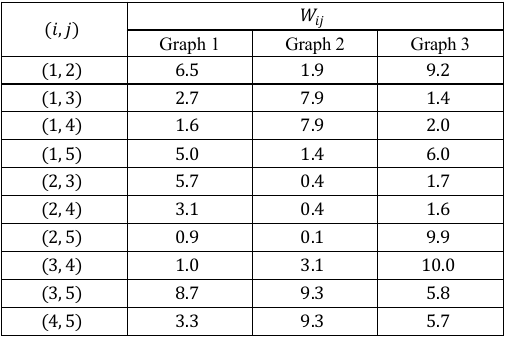}\\
\end{center}
\bigskip

For every graph instance, the five algorithms are repeated ten times. Each algorithm consists of six rotation angles to maintain a constant search space dimension, similarly to the previous example. For all runs of each algorithm, the same initial dataset of three sample points is used, where each sample point consists of six rotation angles. Each run is performed for 50 BO iterations since the algorithms can find the global optimum quickly from 32 possible solutions. The GPR is defined with the QM kernel for uTM-QABOA or the Matérn kernel for all other algorithms. $\nu$ is set to 0.5 for both kernels. All algorithms are performed with the UCB acquisition function with $\alpha$ set to 1. 20,000 simulated annealing steps are performed to maximize the UCB function.

\bigskip
\begin{center}
\includegraphics[width=6.5in,height=2.4in]{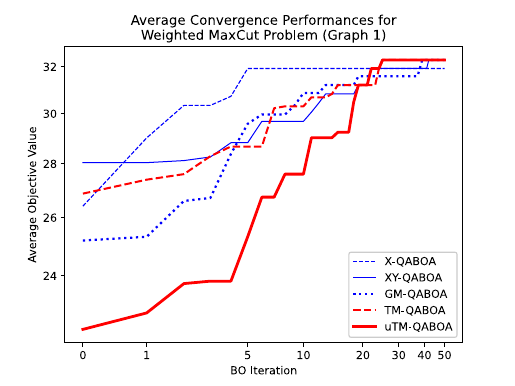}\\
(a)\\
\smallskip
\includegraphics[width=6.5in,height=2.4in]{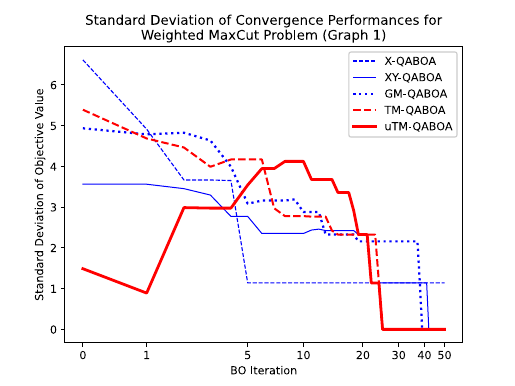}\\
(b)\\
\smallskip
\textbf{FIGURE 4.} (a) Averages and (b) standard deviations for the best
observed objective values for the weighted MaxCut problem with Graph 1.
\end{center}
\bigskip

FIGURE 4, 5, and 6 show the averages and standard deviations of the best observed objective values for Graphs 1, 2, and 3, respectively. For Graphs 1 and 2, TM-QABOA more quickly converges to the global optimum than the single-mixer methods as shown in FIGURE 4(a) and 5(a). For those two graphs, TM-QABOA exhibits lower standard deviation than the single-mixer algorithms as shown in FIGURE 4(b) and 5(b), meaning that the convergence performance of TM-QABOA is more consistent.

\bigskip
\begin{center}
\includegraphics[width=6.5in,height=2.4in]{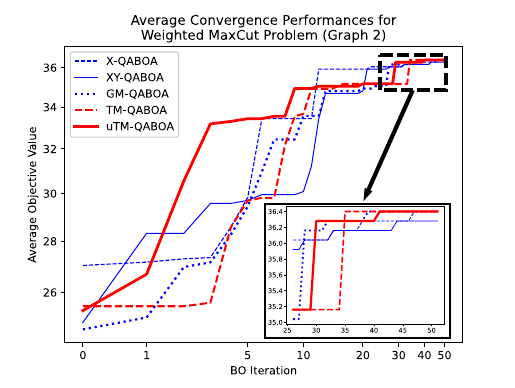}\\
(a)\\
\smallskip
\includegraphics[width=6.5in,height=2.4in]{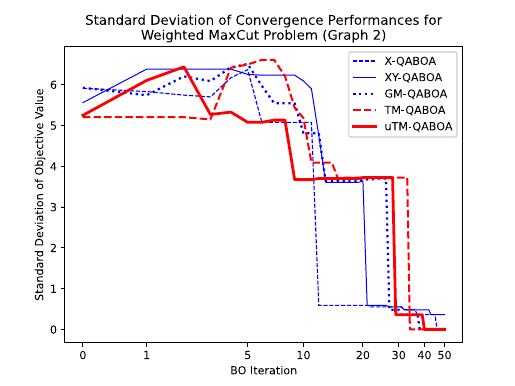}\\
(b)\\
\smallskip
\textbf{FIGURE 5.} (a) Averages and (b) standard deviations for the best
observed objective values for the weighted MaxCut problem with Graph 2.
\end{center}
\bigskip

As shown in FIGURE 4(a), uTM-QABOA achieves a higher objective value than TM-QABOA before converging to the global maximum. Therefore, uTM-QABOA results in the highest searching efficiency for Graph 1. In FIGURE 4(b), uTM-QABOA has the most consistent convergence performance since it has a lower standard deviation than TM-QABOA before converging to the global optimum. For Graph 2, uTM-QABOA is outperformed by GM-QABOA and TM-QABOA. However, it is possible for uTM-QABOA to converge to the global optimum in fewer iterations. At iteration 34, during which TM-QABOA converges to the global optimum, uTM-QABOA results in an average value of 36.28 with a standard deviation of 0.36 in FIGURE 5(a) and (b). In this case, the global optimum is located within one standard deviation of the mean value.

\bigskip
\begin{center}
\includegraphics[width=6.5in,height=2.4in]{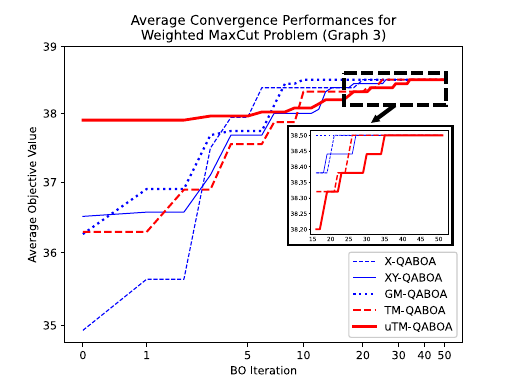}\\
(a)\\
\smallskip
\includegraphics[width=6.5in,height=2.4in]{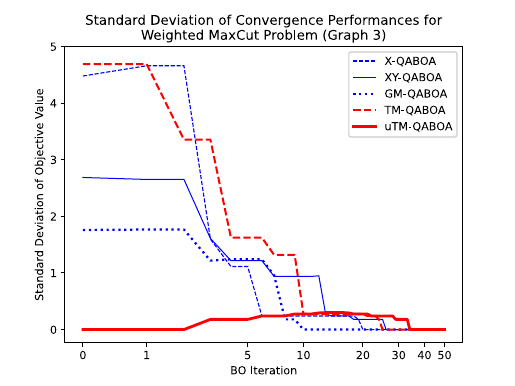}\\
(b)\\
\smallskip
\textbf{FIGURE 6.} (a) Averages and (b) standard deviations for the best
observed objective values for the weighted MaxCut problem with Graph 3.
\end{center}
\bigskip

For Graph 3, TM-QABOA is outperformed by X-QABOA and GM-QABOA, where GM-QABOA converges most quickly to the optimal solution at iteration 10. However, at the same iteration, TM-QABOA results in an average value of 38.32 with a standard deviation of 0.27 in FIGURE 6(a) and (b). Since the global optimum is located within a standard deviation of the mean value, it is possible for TM-QABOA to converge to the global optimum in fewer iterations. The performance of uTM-QABOA is similar to that of TM-QABOA.

It is also observed from all three graphs that GM-QABOA converges to the global optimum more quickly than both X-QABOA and XY-QABOA. Therefore, it is suggested that exploitation through amplitude amplification is important to improve searching efficiency.

\subsection{4.3. Lattice Protein Folding}
\smallskip
\begin{center}
\includegraphics[width=4.7in,height=2.5in]{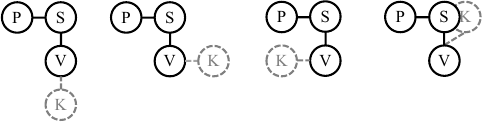}\\
\end{center}
\hspace{0.35in}(a)\hspace{1.1in}(b)\hspace{1.1in}(c)\hspace{1.1in}(d)\\
\begin{center}
\textbf{FIGURE 7.} The PSVK sub-sequence where (a) K is south of V, (b) K is east of V, (c) K is west of V, and (d) K is north of V.
\end{center}
\bigskip

Protein folding is the process where a chain of amino acids is
folded into a three-dimensional protein structure. In the two-dimensional lattice protein folding problem, there are four
choices for the directions of the amino acid bonds: up, down,
left, and right. Suppose a chain consists of six amino acids in
the following order [\ref{perdomo_protein}]: proline (P), serine (S), valine (V),
lysine (K), methionine (M), and alanine (A). The directionality
of the PSVKMA chain is then represented by the binary
sequence $q_aq_bq_cq_dq_eq_fq_gq_hq_iq_j$. The pair $q_aq_b$ represents
the direction from P to S, $q_cq_d$ represents the direction from S
to V, $q_eq_f$ represents the direction from V to K, $q_gq_h$
represents the direction from K to M, and $q_iq_j$
represents the
direction from M to A. Each two-qubit pair has four choices which include 00 (south), 01 (east), 10 (west), and 11 (north).
For instance, suppose that S is east of P and V is south of S.
Then the binary representation of the PSVKMA sequence is
$0100q_1q_2q_3q_4q_5q_6$, where $q_1$
to $q_6$ need to be determined in
the subsequent folding process. This example is to decide the
six remaining folding parameters so as to minimize the energy
of the protein. FIGURE 7 shows the four possible
configurations of the PSVK sub-sequence. The configuration
in FIGURE 7(a) is encoded by $010000q_3q_4q_5q_6$. The
configuration in FIGURE 7(b) is encoded by
$010001q_3q_4q_5q_6$. The configuration in FIGURE 7(c) is
encoded by $010010q_3q_4q_5q_6$. The configuration in FIGURE
7(d) is encoded by $010011q_3q_4q_5q_6$.

\noindent The objective function is

\begin{equation} \label{lattice_obj}
\begin{gathered}
    f(q_{1},q_{2},q_{3},q_{4},q_{5},q_{6}) = - q_{1} + 15q_{1}q_{2} + 4q_{2}q_{3} - 6q_{1}q_{2}q_{3} + 4q_{1}q_{4} \\
    - 15q_{1}q_{2}q_{4} + 15q_{3}q_{4} - 6q_{1}q_{3}q_{4} - 15q_{2}q_{3}q_{4} + 28q_{1}q_{2}q_{3}q_{4} - 4q_{2}q_{5} \\
    + 2q_{1}q_{2}q_{5} + 2q_{2}q_{3}q_{5} + 4q_{1}q_{2}q_{3}q_{5} + 7q_{4}q_{5} + 7q_{5}q_{6} + 2q_{1}q_{4}q_{5} \\
    + 4q_{2}q_{4}q_{5} + 9q_{1}q_{2}q_{4}q_{5} - 20q_{3}q_{4}q_{5} + 4q_{1}q_{3}q_{4}q_{5} + 9q_{2}q_{3}q_{4}q_{5} \\
    - 37q_{1}q_{2}q_{3}q_{4}q_{5} - 4q_{1}q_{6} + 4q_{1}q_{2}q_{6} + 7q_{3}q_{6} + 2q_{1}q_{3}q_{6} + 4q_{2}q_{3}q_{6} \\
    + 9q_{1}q_{2}q_{3}q_{6} + 4q_{1}q_{4}q_{6} - 18q_{3}q_{4}q_{6} + 9q_{1}q_{3}q_{4}q_{6} - 33q_{1}q_{2}q_{3}q_{4}q_{6} \\
    + 2q_{1}q_{5}q_{6} + 4q_{2}q_{5}q_{6} - 20q_{3}q_{5}q_{6} + 9q_{1}q_{2}q_{5}q_{6} + 4q_{1}q_{3}q_{5}q_{6} \\
    + 9q_{2}q_{3}q_{5}q_{6} - 37q_{1}q_{2}q_{3}q_{5}q_{6} - 18q_{4}q_{5}q_{6} + 9q_{1}q_{4}q_{5}q_{6} - 33q_{1}q_{2}q_{4}q_{5}q_{6} \\
    + 53q_{3}q_{4}q_{5}q_{6} - 37q_{1}q_{3}q_{4}q_{5}q_{6} - 33q_{2}q_{3}q_{4}q_{5}q_{6} + 99q_{1}q_{2}q_{3}q_{4}q_{5}q_{6}. \\
\end{gathered}
\end{equation}

The BO search space consists of six rotation angles for the
quantum circuit. Each of the $q_i$’s in Eq. (\ref{lattice_obj}) is implemented
with a Pauli-Z gate as in Eq. (\ref{pauliZsub}). Similar to the previous
example, the GPR models for all five algorithms are initialized
with the same initial dataset of three sample points, where each
sample point corresponds to six values of rotation angles. Each
algorithm is repeated for ten runs, where each run is performed
for 50 BO iterations since the global optimum is easily obtained from 64 possible solutions. The value of $\nu$ is set to 0.5 for both the
Matérn kernel in Eq. (\ref{Matern}) and QM kernel in Eq. (\ref{K_QM}). The
acquisition function is the UCB function in Eq. (\ref{UCB}), where the
value of $\alpha$ is set to 1. 20,000 simulated annealing steps are
performed to find the next sample point.

FIGURE 8(a) and (b) show the averages and standard deviations for the best observed objective values, respectively. In FIGURE 8(a), GM-QABOA, TM-QABOA, and uTM-QABOA converge to the global minimum value of -6 in less than 50 iterations. Among all five algorithms, uTM-QABOA converges to the global optimum in the fewest iterations. It is also observed that GM-QABOA converges to the global optimum in fewer iterations than TM-QABOA. This suggests that the exploitation capability is helpful in improving the searching efficiency. In FIGURE 8(b), uTM-QABOA has the lowest standard deviation after about 5 iterations. This means that uTM-QABOA exhibits the most consistent convergence performance towards the global optimum out of all five algorithms.

\pagebreak
\begin{center}
\includegraphics[width=6.5in,height=2.4in]{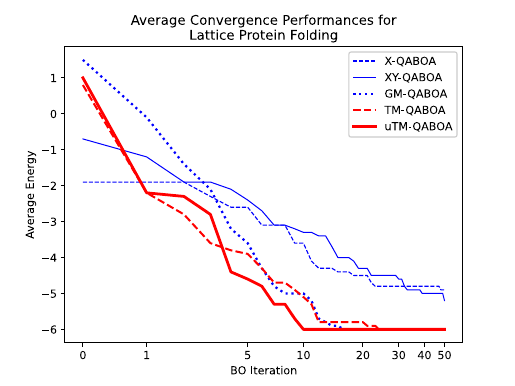}\\
(a)\\
\smallskip
\includegraphics[width=6.5in,height=2.4in]{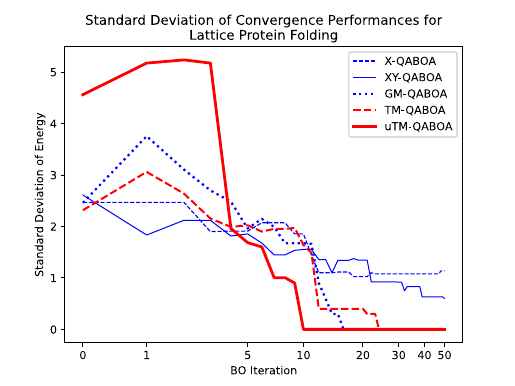}\\
(b)\\
\smallskip
\textbf{FIGURE 8.} (a) Averages and (b) standard deviations for the best
observed energy values for the lattice protein folding problem.
\end{center}
\bigskip

\subsection{4.4. Potential Energy Minimization of the Ionic Helium Hydride Molecule}

The objective of this problem is to minimize the potential
energy of the HeH$^+$ molecule. Two types of variables are
considered. The first type is the bond length $L$ between the
helium and hydrogen atoms, which continuously ranges from 0.1 to 3 Å. The second type is the orbital configuration, which
is represented by binary variables.

The orbital configuration of HeH$^+$
can be fully represented
with four qubits, each of which corresponds to a spin orbital.
For a molecular Hamiltonian, the number of qubits can be
reduced with the qubit tapering technique [\ref{setia_qubitreduce}]. In qubit tapering, those Pauli operators which commute with the molecular Hamiltonian are determined. These Pauli operators encode the symmetries of the Hamiltonian. A new Hamiltonian with a smaller size is computed from these Pauli operators and used in computing the energy instead of the original Hamiltonian. For HeH$^+$, qubit tapering reduces the
number of qubits from four to two. Therefore, two binary
variables represent the orbital configuration of the molecule.

\bigskip
\begin{center}
\includegraphics[width=6.5in,height=2.4in]{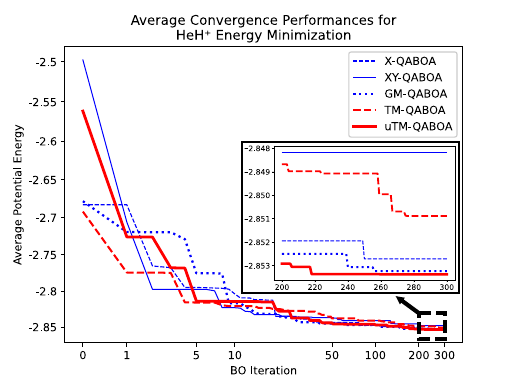}\\
(a)\\
\smallskip
\includegraphics[width=6.5in,height=2.4in]{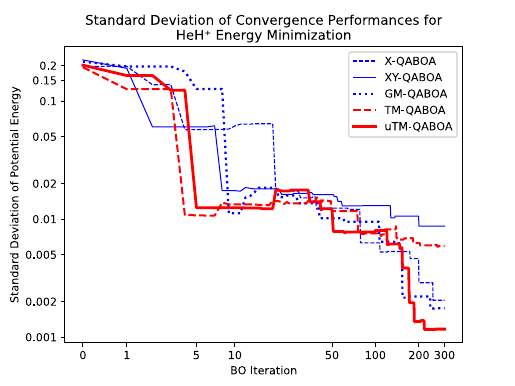}\\
(b)\\
\smallskip
\textbf{FIGURE 9.} (a) Averages and (b) standard deviations for the best
observed potential energy values for the HeH$^+$ energy minimization problem.
\end{center}
\bigskip

Each of the five algorithms is repeated for ten runs, where
each run is performed for 300 BO iterations. Compared to the previous three examples, the number of BO iterations is much larger. It is expected that finding the global optimum for continuous variables is more difficult. The initial dataset for all runs consists of the same five sample points randomly sampled from the search space. Each sample point corresponds to the six values of rotation angles and one value of $L$. The phase-separating Hamiltonian is the qubit-tapered
molecular Hamiltonian computed with the PennyLane Python
library. The quantum circuit is performed on two qubits which
represent the orbital configuration of HeH$^+$
. The value of $\nu$ is
0.5 for both the Matérn and QM kernels. The acquisition
function is the UCB function with $\alpha$ set to 1, where this
function is maximized with 20,000 simulated annealing steps.

FIGURE 9(a) and (b) show the averages and standard deviations for the best observed objective values, respectively. The closer view in FIGURE 9(a) shows the convergence during the last 100 iterations. Overall, uTM-QABOA exhibits the best optimization performance. In FIGURE 9(a), uTM-QABOA results in the lowest average objective value after 300 iterations. This is a noticeable improvement over TM-QABOA, which converges to a higher average objective value than X-QABOA and GM-QABOA. GM-QABOA exhibits the second highest searching efficiency, which suggests the importance of exploitation in improving the searching efficiency. In FIGURE 9(b), uTM-QABOA has the lowest standard deviation after about 200 iterations.

\subsection{4.5. Welded Beam Design}

FIGURE 10 illustrates a welded beam [\ref{tran_mixedint}] subjected to a
downward point load of $F$ = 6,000 lb at the free end. The
beam length is known to be $L$ = 14 in. The continuous
design variables $\textbf{x}_c$ include the weld thickness $h$, the welded joint
length $l$, the beam width $t$, and the beam thickness $b$. Two
discrete design variables include the weld type $w$ and the bulk
material type $m$.

\begin{center}
\includegraphics[width=4in,height=2.4in]{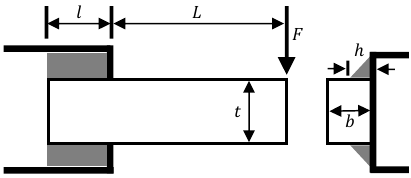}\\
\smallskip
\textbf{FIGURE 10.} Welded beam design problem.
\end{center}
\bigskip
\pagebreak

The objective of the problem is to minimize the cost of the
welded beam, defined as

\begin{equation} \label{beam_obj}
    f(w,m,h,l,t,b) = (1 + C_1(m))(wt + l)h^2 + C_2(m)tb(L + l)
\end{equation}

\noindent where $C_1(m)$ and $C_2(m)$ are the costs per volume of the
welded material and bar stock, respectively. $C_1$ and $C_2$
depend on material type $m$. The design variables are
subjected to a buckling load constraint defined as

\begin{equation} \label{beam_constraint}
    b - h \geq 0.
\end{equation}

It is noted that this problem is a simplified version of the
original welded beam design problem [\ref{datta_discrete}], where additional
constraints are imposed.

For the continuous variables, 0.0625 in $\leq h \leq$ 2 in,
0.1 in $\leq l \leq$ 10 in, 2 in $\leq t \leq$ 20 in, and 0.0625 in $\leq b \leq$ 2 in. The discrete variable $w$ is set to 0 or 1 for two-sided and
four-sided welding, respectively. The discrete variable $m$ is set
to 1, 2, 3, or 4 which corresponds to steel, cast iron, aluminum,
and brass, respectively. When $m$ = 1, $C_1$ = \$0.1047 in$^{-3}$
and  $C_2$ = \$0.0481 in$^{-3}$. When $m$ = 2, $C_1$ = \$0.0489 in$^{-3}$
and $C_2$ = \$0.0224 in$^{-3}$. When $m$ = 3, $C_1$ = \$0.5235 in$^{-3}$
and $C_2$ = \$0.2405 in$^{-3}$. When $m$ = 4, $C_1$ = \$0.5584 in$^{-3}$
and $C_2$ = \$0.2566 in$^{-3}$.

Since there are two choices for $w$ and four choices for $m$,
the optimization problem is solved with a three-qubit system.
The objective function in Eq. (\ref{beam_obj}) can be reformulated as the
function in Eq. (\ref{mod_beam_obj}) in Appendix, where $q_1$
is a binary variable
which corresponds to values of $w$, and $q_2$ and $q_3$ are binary
variables which correspond to values of $m$.

Each of the five algorithms is repeated for ten runs, where
each run is performed for 300 BO iterations. The initial dataset
is constructed with ten sample points randomly acquired from
the search space. Each sample point corresponds to the six
values of rotation angles, and four values for the continuous
variables. The value of $\nu$ is set to 0.5 for both the Matérn and
QM kernels. The UCB acquisition function, with $\alpha$ set to 1, is
maximized with 20,000 simulated annealing steps. If the
buckling load constraint is violated at a sample point, the UCB
function value is set to -100,000 to prevent the sample point
from being selected.

Out of all the tested algorithms, uTM-QABOA exhibits the best optimization performance. As shown in FIGURE 11(a), uTM-QABOA results in the lowest average objective value after 300 iterations. It also has the lowest standard deviation after about 100 iterations, as shown in FIGURE 11(b).

\bigskip
\begin{center}
\includegraphics[width=6.5in,height=2.4in]{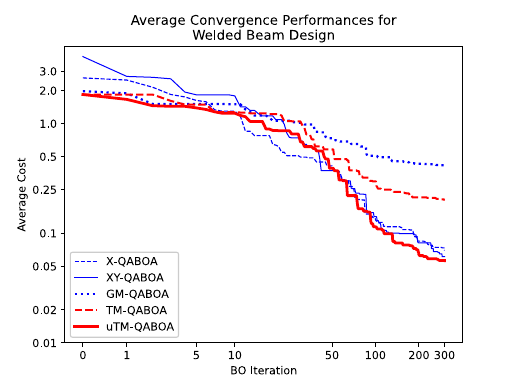}\\
(a)\\
\smallskip
\includegraphics[width=6.5in,height=2.4in]{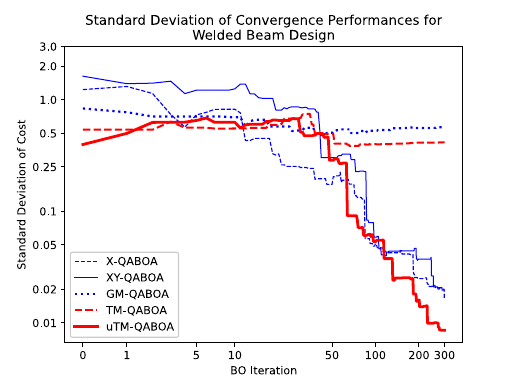}\\
(b)\\
\smallskip
\textbf{FIGURE 11.} (a) Averages and (b) standard deviations for the best
observed cost values for the welded beam design problem.
\end{center}

Meanwhile, after 300 iterations, TM-QABOA results in a higher average objective value than X-QABOA, XY-QABOA, and uTM-QABOA. Nevertheless, the standard deviation of TM-QABOA is also higher than those of the other three algorithms. This indicates that the performance of TM-QABOA is not consistent, although it may sometimes result in better objective values. The performance of GM-QABOA is similar to that of TM-QABOA, which again shows the importance of exploitation consideration.
\pagebreak

\subsection{4.6. Speed Reducer Design}
The objective of this problem is to minimize the weight of a
speed reducer, defined as [\ref{tran_mixedint}]

\begin{equation} \label{reducer_obj}
\begin{gathered}
    f(x_1, x_2, x_3, x_4, x_5, x_6, x_7) = 0.7854x_1x_2^2(3.3333x_3^2 + 14.9334x_3\\
    - 43.0934) - 1.508x_1(x_6^2 + x_7^2) + 7.4777(x_6^3 + x_7^3) \\
    + 0.7854(x_4x_6^2 + x_5x_7^2),
\end{gathered}
\end{equation}

where $x_1$, $x_2$, $x_3$, $x_4$, $x_5$, $x_6$, and $x_7$ are the face weight,
module of teeth, number of pinion teeth, length of the first
shaft between bearings, length of the second shaft between
bearings, first shaft diameter, and second shaft diameter,
respectively. $x_3$
is an integer-valued variable with 16 choices
ranging from 15 to 30. The continuous variables $\textbf{x}_c$ include $x_1, x_2, x_4, x_5, x_6,$ and $x_7$. The ranges for the continuous
variables are 2.6 $\leq x_1 \leq$ 3.6, 0.7 $\leq x_2 \leq$ 0.8, 7.3 $\leq x_4 \leq$
8.3, 7.8 $\leq x_5 \leq$ 8.3, 2.9 $\leq x_6 \leq$ 3.9, and 5 $\leq x_7 \leq$ 5.5. In
this work, the constraints of the original speed reducer design
problem [\ref{cagnina_eng}] are ignored.

The problem is solved with four qubits since there are 16
choices for $x_3$. The objective function in Eq. (\ref{reducer_obj}) can be
reformulated as the function in Eq. (\ref{mod_reducer_obj}) in Appendix, where $q_1$, $q_2$, $q_3$, and $q_4$ are binary variables which correspond to
values of $x_3$.

Each of the five algorithms is repeated for ten runs, where
each run is performed for 300 BO iterations. The initial dataset
is constructed with 10 sample points randomly acquired from
the search space. Each sample point corresponds to the six
values of rotation angles, and six values of the continuous
variables. The value of $\nu$ for the Matérn and QM kernels is set
to 0.5. The UCB acquisition function, with $\alpha$ set to 1, is
maximized with 20,000 simulated annealing steps.

As shown in FIGURE 12(a), uTM-QABOA converges to the lowest average objective value after 300 iterations. In FIGURE 12(b), uTM-QABOA results in the lowest standard deviation after about 100 iterations. There are noticeable improvements in both average and standard deviation over TM-QABOA. It is also observed that GM-QABOA results in the second lowest average objective value after 300 iterations. This also suggests that exploitation is important in finding the optimal solution efficiently.

\pagebreak
\begin{center}
\includegraphics[width=6.5in,height=2.4in]{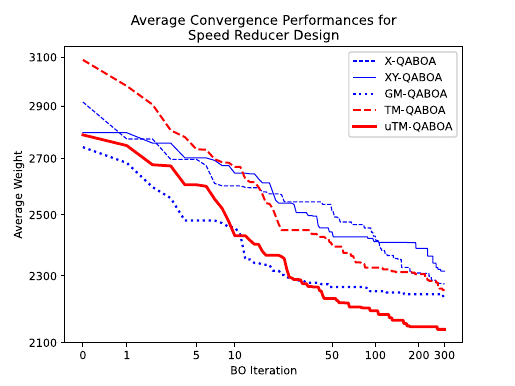}\\
(a)\\
\smallskip
\includegraphics[width=6.5in,height=2.4in]{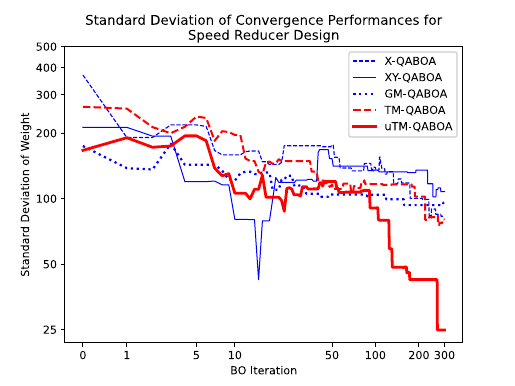}\\
(b)\\
\smallskip
\textbf{FIGURE 12.} (a) Averages and (b) standard deviations for the best
observed weight values for the speed reducer design problem.
\end{center}
\bigskip

\subsection{4.7. Pressure Vessel Design}
The objective of this problem is to minimize the cost of a cylindrical pressure vessel, defined as [\ref{tran_mixedint}]

\begin{equation} \label{pressure_obj}
    f(x_1,x_2,x_3,x_4) = 0.6224x_1x_3x_4 + 1.7781x_2x_3^2 + 3.1661x_1^2x_4 + 19.84x_1^2x_3,
\end{equation}

\noindent where $x_1$
is the hemisphere thickness, $x_2$
is the cylindrical
shell thickness, $x_3$
is the hemisphere inner radius, and $x_4$
is the
cylinder length. The design variables are subjected to a volume
constraint defined as

\begin{equation} \label{pressure_constraint}
    -{\pi}x_3^2x_4^2 - \frac{4}{3}x_3^3 + 1296000 \leq 0.
\end{equation}

\noindent In this problem, $x_1$ and $x_2$ are integer-valued variables each
ranging from 3 to 6. The continuous variables $\textbf{x}_c$ include $x_3$ and $x_4$ which each range from 10 to 150. The variable values
are set so that all constraints of the original pressure vessel
design problem [\ref{cagnina_eng}] are satisfied. Since there are 16 possible
combinations of $x_1$ and $x_2$, four qubits are needed to solve the
problem. The objective function defined in Eq. (\ref{pressure_obj}) can be
reformulated as the function in Eq. (\ref{mod_pressure_obj}) in Appendix, where
$q_1$ and $q_2$ correspond to values of $x_1$, and $q_3$ and $q_4$
correspond to values of $x_2$.

Each of the five algorithms is repeated for ten runs, where
each run is performed for 300 BO iterations. The initial dataset
is constructed with ten random sample points. Each sample
point corresponds to the six values of rotation angles, and two
values of continuous variables. The value of $\nu$ is set to 0.5 for
both the Matérn and QM kernels. The UCB acquisition
function, with $\alpha$ set to 1, is maximized with 20,000 simulated
annealing steps. If the volume constraint is violated for a sample point, the UCB function value is set to $-$10,000,000 to
prevent the sample point from being selected.

FIGURE 13(a) and (b) show the averages and standard deviations of the best observed objective values, respectively. The closer view in FIGURE 13(a) shows the convergence during the last 100 iterations. The averages from all five algorithms are very similar. uTM-QABOA has the highest value. Nevertheless, the difference between the objective values of uTM-QABOA and GM-QABOA is 111.5. The standard deviation of the best objective values of uTM-QABOA is 108.0. With the large standard deviation, uTM-QABOA may still obtain the lowest objective value.

In this example, GM-QABOA converges to the lowest objective value on average with the smallest standard deviation. It has the best performance out of the five algorithms. Similar to the previous five examples, amplitude amplification has shown the benefit of improving the searching efficiency, where solutions with better objective values are more favorably selected. The performance of TM-QABOA is similar to that of GM-QABOA. In this case, TM-QABOA converges to the second lowest objective value on average with a slightly higher standard deviation than GM-QABOA. The results of those two algorithms demonstrate the usefulness of exploitation and the importance of dynamic exploration-exploitation balance in achieving the optimal solution efficiently.

\pagebreak
\begin{center}
\includegraphics[width=6.5in,height=2.4in]{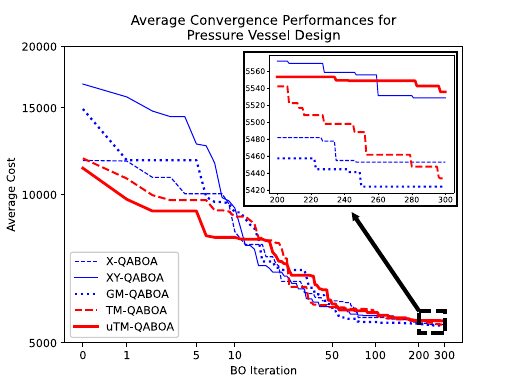}\\
(a)\\
\smallskip
\includegraphics[width=6.5in,height=2.4in]{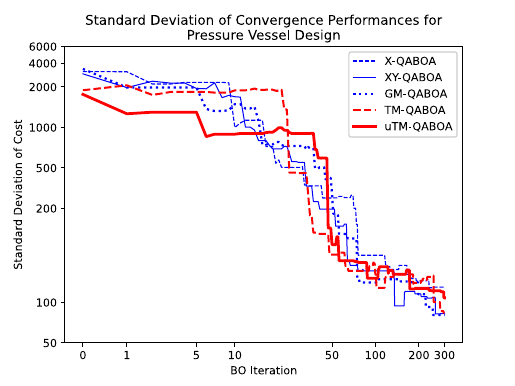}\\
(b)\\
\smallskip
\textbf{FIGURE 13.} (a) Averages and (b) standard deviations for the best
observed cost values for the pressure vessel design problem.
\end{center}
\bigskip

\section{5. Concluding Remarks}
In this paper, two versions of QAOA are proposed, where surrogate-based Bayesian optimization is used as the classical optimizer. They are named as TM-QABOA and uTM-QABOA. In TM-QABOA, the quantum circuit consists of two mixers to improve the exploration-exploitation balance. The Pauli-gate mixer is for exploration, while the generalized Grover mixer is for exploitation. In uTM-QABOA, a new QM kernel is proposed to consider the variation of objective values in the sampling process. The formulation based on the kurtosis allows for the dynamic control of the variation based on the optimality of the objective values. As a result, this increases the chance of obtaining the optimum, thus improving the searching efficiency.

The proposed TM-QABOA and uTM-QABOA can be applied to solve mixed-integer problems. The rotation angles and continuous variables are optimized with Bayesian optimization, while the discrete variables are optimized through the quantum circuit. The new algorithms are tested with nine problems. They are compared with three single-mixer QABOA's, where the same Bayesian optimization framework is used as the classical optimizer. The results show that the proposed uTM-QABOA has the best performance in efficiency and consistency for five out of the nine tested problems. The results also show that GM-QABOA with the generalized Grover mixer performs the best among the three single-mixer algorithms, thereby demonstrating the benefit of exploitation in improving searching efficiency.

The purpose of TM-QABOA and uTM-QABOA is to improve the searching efficiency of QAOA’s. QAOA’s suffer from low searching efficiency when the circuit depth increases, since a large number of rotation angles need to be optimized. This is known as curse of dimensionality for optimization. The surrogate-based Bayesian optimization can improve the searching efficiency with the guidance of the GPR model. The results in this paper show the advantage of the proposed QABOA’s. However, all nine test problems involve no more than seven variables and six qubits. For a small number of basis states, the performances of single mixers are reasonable. The Pauli-X mixer can drastically alter the amplitudes among a few basis states. As the number of basis states increases, the number of states with non-zero amplitudes increases and it is more difficult for the generalized Grover operator to increase the target basis amplitude to 1. For problems with larger numbers of qubits, the performance of the mixers could become worse. To fully understand the performance of QABOA’s, larger problems with more qubits need to be studied in the future. In addition, TM-QABOA and uTM-QABOA are tested at low circuit depths in this paper. Sensitivity studies are also needed to evaluate the effect of circuit depths.

Similar to other quantum optimization algorithms, uTM-QABOA involves repeated sampling of the same quantum circuit to obtain a more reliable estimation of the optimum. The sampling is also used to approximate the basis state distributions in order to estimate the kurtosis in the QM kernel. To further improve the computational efficiency of uTM-QABOA, better sampling strategies are needed. For instance, variational inference can be applied where the basis state distributions are parametrized, and a smaller number of samples are needed to train and optimize those parameters.

\section*{Appendix}
The modified objective function for the welded beam design problem is

\begin{equation} \label{mod_beam_obj}
\begin{gathered}
    g\left( q_{1},q_{2},q_{3},h,l,t,b \right) = \left( 1 - q_{1} \right)\left( 1 - q_{2} \right)\left( 1 - q_{3} \right)f(0,1,h,l,t,b) \\
    + \left( 1 - q_{1} \right)\left( 1 - q_{2} \right)q_{3}f(0,2,h,l,t,b) + \left( 1 - q_{1} \right)q_{2}\left( 1 - q_{3} \right)f(0,3,h,l,t,b) \\
    + \left( 1 - q_{1} \right)q_{2}q_{3}f(0,4,h,l,t,b) + q_{1}\left( 1 - q_{2} \right)\left( 1 - q_{3} \right)f(1,1,h,l,t,b) \\
    + q_{1}\left( 1 - q_{2} \right)q_{3}f(1,2,h,l,t,b) + q_{1}q_{2}\left( 1 - q_{3} \right)f(1,3,h,l,t,b) \\
    + q_{1}q_{2}q_{3}f(1,4,h,l,t,b).
\end{gathered}
\end{equation}

\noindent The modified objective function for the speed reducer design problem is

\begin{equation} \label{mod_reducer_obj}
\begin{gathered}
    g\left( q_{1},q_{2},q_{3},q_{4},x_{1},x_{2},x_{4},x_{5},x_{6},x_{7} \right) \\
    = \left( 1 - q_{1} \right)\left( 1 - q_{2} \right)\left( 1 - q_{3} \right)(1 - q_{4})f\left( x_{1},x_{2},15,x_{4},x_{5},x_{6},x_{7} \right) \\
    + \left( 1 - q_{1} \right)\left( 1 - q_{2} \right)\left( 1 - q_{3} \right)q_{4}f\left( x_{1},x_{2},16,x_{4},x_{5},x_{6},x_{7} \right) \\
    + \left( 1 - q_{1} \right)\left( 1 - q_{2} \right)q_{3}\left( 1 - q_{4} \right)f\left( x_{1},x_{2},17,x_{4},x_{5},x_{6},x_{7} \right) \\
    + \left( 1 - q_{1} \right)\left( 1 - q_{2} \right)q_{3}q_{4}f\left( x_{1},x_{2},18,x_{4},x_{5},x_{6},x_{7} \right) \\
    + \left( 1 - q_{1} \right)q_{2}\left( 1 - q_{3} \right)\left( 1 - q_{4} \right)f\left( x_{1},x_{2},19,x_{4},x_{5},x_{6},x_{7} \right) \\
    + \left( 1 - q_{1} \right)q_{2}\left( 1 - q_{3} \right)q_{4}f\left( x_{1},x_{2},20,x_{4},x_{5},x_{6},x_{7} \right) \\
    + \left( 1 - q_{1} \right)q_{2}q_{3}\left( 1 - q_{4} \right)f\left( x_{1},x_{2},21,x_{4},x_{5},x_{6},x_{7} \right) \\
    + \left( 1 - q_{1} \right)q_{2}q_{3}q_{4}f\left( x_{1},x_{2},22,x_{4},x_{5},x_{6},x_{7} \right) \\
    + q_{1}\left( 1 - q_{2} \right)\left( 1 - q_{3} \right)\left( 1 - q_{4} \right)f\left( x_{1},x_{2},23,x_{4},x_{5},x_{6},x_{7} \right) \\
    + q_{1}\left( 1 - q_{2} \right)\left( 1 - q_{3} \right)q_{4}f\left( x_{1},x_{2},24,x_{4},x_{5},x_{6},x_{7} \right) \\
    + q_{1}\left( 1 - q_{2} \right)q_{3}\left( 1 - q_{4} \right)f\left( x_{1},x_{2},25,x_{4},x_{5},x_{6},x_{7} \right) \\
    + q_{1}\left( 1 - q_{2} \right)q_{3}q_{4}f\left( x_{1},x_{2},26,x_{4},x_{5},x_{6},x_{7} \right) \\
    + q_{1}q_{2}\left( 1 - q_{3} \right)\left( 1 - q_{4} \right)f\left( x_{1},x_{2},27,x_{4},x_{5},x_{6},x_{7} \right) \\
    + q_{1}q_{2}\left( 1 - q_{3} \right)q_{4}f\left( x_{1},x_{2},28,x_{4},x_{5},x_{6},x_{7} \right) \\
    + q_{1}q_{2}q_{3}\left( 1 - q_{4} \right)f\left( x_{1},x_{2},29,x_{4},x_{5},x_{6},x_{7} \right) \\
    + q_{1}q_{2}q_{3}q_{4}f\left( x_{1},x_{2},30,x_{4},x_{5},x_{6},x_{7} \right) 
\end{gathered}
\end{equation}

\pagebreak
\noindent The modified objective function for the pressure vessel design problem is

\begin{equation} \label{mod_pressure_obj}
\begin{gathered}
    g\left( q_{1},q_{2},q_{3},q_{4},x_{3},x_{4} \right) = \left( 1 - q_{1} \right)\left( 1 - q_{2} \right)\left( 1 - q_{3} \right)(1 - q_{4})f\left( 3, 3,x_{3},x_{4} \right) \\
    + \left( 1 - q_{1} \right)\left( 1 - q_{2} \right)\left( 1 - q_{3} \right)q_{4}f\left( 3, 4,x_{3},x_{4} \right) \\
    + \left( 1 - q_{1} \right)\left( 1 - q_{2} \right)q_{3}\left( 1 - q_{4} \right)f\left( 3, 5,x_{3},x_{4} \right) \\
    + \left( 1 - q_{1} \right)\left( 1 - q_{2} \right)q_{3}q_{4}f\left( 3, 6,x_{3},x_{4} \right) \\
    + \left( 1 - q_{1} \right)q_{2}\left( 1 - q_{3} \right)\left( 1 - q_{4} \right)f\left( 4, 3,x_{3},x_{4} \right) \\
    + \left( 1 - q_{1} \right)q_{2}\left( 1 - q_{3} \right)q_{4}f\left( 4, 4,x_{3},x_{4} \right) \\
    + \left( 1 - q_{1} \right)q_{2}q_{3}\left( 1 - q_{4} \right)f\left( 4, 5,x_{3},x_{4} \right) \\
    + \left( 1 - q_{1} \right)q_{2}q_{3}q_{4}f\left( 4, 6,x_{3},x_{4} \right) \\
    + q_{1}\left( 1 - q_{2} \right)\left( 1 - q_{3} \right)\left( 1 - q_{4} \right)f\left( 5, 3,x_{3},x_{4} \right) \\
    + q_{1}\left( 1 - q_{2} \right)\left( 1 - q_{3} \right)q_{4}f\left( 5, 4,x_{3},x_{4} \right) \\
    + q_{1}\left( 1 - q_{2} \right)q_{3}\left( 1 - q_{4} \right)f\left( 5, 5,x_{3},x_{4} \right) \\
    + q_{1}\left( 1 - q_{2} \right)q_{3}q_{4}f\left( 5, 6,x_{3},x_{4} \right) + q_{1}q_{2}\left( 1 - q_{3} \right)\left( 1 - q_{4} \right)f\left( 6, 3,x_{3},x_{4} \right) \\
    + q_{1}q_{2}\left( 1 - q_{3} \right)q_{4}f\left( 6, 4,x_{3},x_{4} \right) + q_{1}q_{2}q_{3}\left( 1 - q_{4} \right)f\left( 6, 5,x_{3},x_{4} \right) \\
    + q_{1}q_{2}q_{3}q_{4}f\left( 6, 6,x_{3},x_{4} \right).
\end{gathered}
\end{equation}

\section*{References}
\renewcommand*\labelenumi{[\theenumi]}
\begin{enumerate}
\item \label{wang_QCOverview} Y. Wang, J. E. Kim, and K. Suresh, “Opportunities and challenges of quantum computing for engineering optimization,” Journal of Computing and Information Science in Engineering, vol. 23, no. 6, Aug. 2023, DOI. \href{https://doi.org/10.1115/1.4062969}{10.1115/1.4062969}
\item \label{wang_QuantumWalk} Y. Wang, “Global optimization with quantum walk enhanced Grover search,”, in 2014 ASME International Design Engineering Technical Conferences \& The Computer and Information in Engineering Conference, Buffalo, NY, USA, 2014, DOI. \href{https://doi.org/10.1115/DETC2014-34634}{10.1115/DETC2014-34634}.
\item \label{wang_StochasticDiff} Y. Wang, “Simulating stochastic diffusions by quantum walks,” in 2013 ASME International Design Engineering Technical Conferences \& The Computer and Information in Engineering Conference, Portland OR, USA, 2013, DOI. \href{https://doi.org/10.1115/DETC2013-12739}{10.1115/DETC2013-12739}.
\item \label{wang_StochasticDynamics} Y. Wang, "Accelerating stochastic dynamics simulation with continuous-time quantum walks," in 2016 ASME International Design Engineering Technical Conferences \& The Computer and Information in Engineering Conference, Charlotte, NC, USA, 2016, DOI. \href{https://doi.org/10.1115/DETC2016-59420}{10.1115/DETC2016-59420}.
\item \label{farhi_QAOA}
  E. Farhi, J. Goldstone, and S. Gutmann, ``A quantum approximate
  optimization algorithm,'' 2014, \emph{arXiv preprint arXiv:1411.4028}. 
\item \label{zhou_QAOA}
  L. Zhou, S. T. Wang, S. Choi, H. Pichler, and M. D. Lukin, ``Quantum
  approximate optimization algorithm: performance, mechanism, and
  implementation on near-term devices,'' \emph{Phys. Rev. X}, vol. 10,
  no. 2, pp. 021067, Jun. 2020, DOI.
\href{https://doi.org/10.1103/PhysRevX.10.021067}{10.1103/PhysRevX.10.021067}.
\item \label{wang_QAOAmaxcut}
  Z. Wang, S. Hadfield, Z. Jiang, and E. G. Rieffel, ``Quantum
  approximate optimization algorithm for MaxCut: a fermionic view,''
  \emph{Phys. Rev. A}, vol. 97, no. 2, pp. 022304, Feb. 2018, DOI.
  \href{https://doi.org/10.1103/PhysRevA.97.022304}{10.1103/PhysRevA.97.022304}.
\item \label{shaydulin_QAOA}
  R. Shaydulin and S. M. Wild, ``Exploiting symmetry reduces the cost of
  training QAOA,'' \emph{IEEE Transactions on Quantum Engineering}, vol.
  2, pp. 1--9, Mar. 2021, DOI.
  \href{https://doi.org/10.1109/TQE.2021.3066275}{10.1109/TQE.2021.3066275}.
\item \label{tate_warmQAOA}
  R. Tate, M. Farhadi, C. Herold, G. Mohler, and S. Gupta, ``Bridging
  classical and quantum with SDP initialized warm-starts for QAOA,''
  \emph{ACM Transactions of Quantum Computing}, vol. 4, no. 2, pp.
  1--39, Feb. 2023, DOI.
  \href{https://doi.org/10.1145/3549554}{10.1145/3549554}.
\item \label{cook_QAOAz}
  J. Cook, S. Eidenbenz, and A. Bärtschi, ``The quantum alternating
  operator ansatz on maximum k-vertex cover,'' in \emph{2020 IEEE
  International Conference on Quantum Computing and Engineering (QCE)},
  Denver, CO, USA, 2020, pp. 83--92. DOI.
  \href{https://doi.org/10.1109/QCE49297.2020.00021}{10.1109/QCE49297.2020.00021}.
\item \label{campbell_QAOA}
  C. Campbell and E. Dahl, ``QAOA of the highest order,'' in \emph{2022
  IEEE 19th International Conference on Software Architecture Companion
  (ICSA-C)}, Honolulu, HI, USA, 2022, pp. 141--146. DOI.
  \href{https://doi.org/10.1109/ICSA-C54293.2022.00035}{10.1109/ICSA-C54293.2022.00035}.
\item \label{hadfield_QAOAz}
  S. Hadfield, Z. Wang, B. O'Gorman, E. G. Rieffel, D. Venturelli, and
  R. Biswas, ``From the quantum approximate optimization algorithm to a
  quantum alternating operator ansatz,'' \emph{Algorithms}, vol. 12, no.
  2, pp. 34, Feb. 2019, DOI.
  \href{https://doi.org/10.3390/a12020034}{{10.3390/a12020034}}.
\item \label{wang_XY}
  Z. Wang, N. C. Rubin, J. M. Dominy, and E. G. Rieffel, ``XY mixers:
  analytical and numerical results for the quantum alternating operator
  ansatz,'' \emph{Phys Rev. A}, vol. 101, no. 1, pp. 012320, Jan. 2020,
  DOI.
  \href{https://doi.org/10.1103/PhysRevA.101.012320}{10.1103/PhysRevA.101.012320}.
\item \label{marsh_quantum}
  S. Marsh and J. B. Wang, ``A quantum walk-assisted approximate
  algorithm for bounded NP optimisation problems,'' \emph{Quantum
  Information Processing}, vol. 18, pp. 1--18, Jan. 2019, DOI.
  \href{https://doi.org/10.1007/s11128-019-2171-3}{10.1007/s11128-019-2171-3}.
\item \label{bartschi_grover}
  A. Bärtschi and S. Eidenbenz, ``Grover mixers for QAOA: shifting
  complexity from mixer design to state preparation,'' in \emph{2020
  IEEE International Conference on Quantum Computing and Engineering
  (QCE)}, Denver, CO, USA, 2020, pp. 72--82. DOI.
  \href{https://doi.org/10.1109/QCE49297.2020.00020}{{10.1109/QCE49297.2020.00020}}.
\item \label{ywang_QABOA}
  Y. Wang, ``A quantum approximate Bayesian optimization algorithm for
  continuous problems,'' in \emph{2021 IISE Annual Conference \& Expo},
  New Orleans, LA, USA, 2021, pp. 235--240.
\item \label{govia_freedom}
  L. C. G. Govia, C. Poole, M. Saffman, and H. K. Krovi, ``Freedom of
  the mixer rotation axis improves performance in the quantum
  approximate optimization algorithm,'' \emph{Phys Rev. A}, vol. 104,
  no. 6, pp. 062428, Dec. 2021, DOI.
  \href{https://doi.org/10.1103/PhysRevA.104.062428}{10.1103/PhysRevA.104.062428}.
\item \label{chen_entangle}
  Y. Chen, L. Zhu, N. J. Mayhall, E. Barnes, and S. E. Economou, ``How
  much entanglement do quantum optimization algorithms require?'' in
  \emph{Quantum 2.0 Conference 2022}, Boston, MA, USA, 2022, pp. QM4A-2.
  DOI.
  \href{https://doi.org/10.1364/QUANTUM.2022.QM4A.2}{10.1364/QUANTUM.2022.QM4A.2}.
\item \label{brandhofer_QAOA}
  S. Brandhofer \emph{et al.}, ``Benchmarking the performance of
  portfolio optimization with QAOA,'' \emph{Quantum Information
  Processing}, vol. 22, no. 1, pp. 1--27, Dec. 2022, DOI.
  \href{https://doi.org/10.1007/s11128-022-03766-5}{10.1007/s11128-022-03766-5}.
\item \label{niroula_constrained}
  P. Niroula \emph{et al.}, ``Constrained quantum optimization for
  extractive summarization on a trapped-ion quantum computer,''
  \emph{Scientific Reports}, vol. 12, no. 1, pp. 17171, Oct. 2022, DOI.
  \href{https://doi.org/10.5281/zenodo.6819861}{{10.5281/zenodo.6819861}}.
\item \label{larose_mixerphaser}
  R. LaRose, E. Rieffel, and D. Venturelli, "Mixer-phaser ansätze for
  quantum optimization with hard constraints,'' \emph{Quantum Machine
  Intelligence}, vol. 4, no. 17, Jun. 2022, DOI.
  \href{https://doi.org/10.1007/s42484-022-00085-x}{10.1007/s42484-022-00085-x}.
\item \label{grover_database}
  L. K. Grover, ``A fast quantum mechanical algorithm for database
  search,'' in \emph{STOC '96: Proc. of the Twenty-Eighth Annual ACM
  Symposium on Theory of Computing}, Philadelphia, PA, USA, 1996, pp.
  212--219. DOI.
  \href{https://doi.org/10.1145/237814.237866}{{10.1145/237814.237866}}.
\item \label{yu_QAOAbias}
  Y. Yu, C. Cao, C. Dewey, X. B. Wang, N. Shannon, and R. Joynt,
  ``Quantum approximate optimization algorithm with adaptive bias
  fields,'' \emph{Phys. Rev. Research}, vol. 4, no. 2, pp. 023249, Jun.
  2022, DOI.
  \href{https://doi.org/10.1103/PhysRevResearch.4.023249}{10.1103/PhysRevResearch.4.023249}.
\item \label{chancellor_domain}
  N. Chancellor, ``Domain wall encoding of discrete variables for
  quantum annealing and QAOA,'' \emph{Quantum Science and Technology},
  vol. 4, no. 4, pp. 045004, Aug. 2019, DOI.
  \href{https://doi.org/10.1088/2058-9565/ab33c2}{10.1088/2058-9565/ab33c2}.
\item \label{xue_QAOAnoise}
  C. Xue, Z. Y. Chen, Y. C. Wu, and G. P. Guo, ``Effects of quantum
  noise on quantum approximate optimization algorithm,'' \emph{Chinese
  Physics Letters}, vol. 38, no. 3, pp. 030302, Mar. 2021, DOI.
  \href{https://doi.org/10.1088/0256-307X/38/3/030302}{10.1088/0256-307X/38/3/030302}.
\item \label{marshall_QAOAnoise}
  J. Marshall, F. Wudarski, S. Hadfield, and T. Hogg, ``Characterizing
  local noise in QAOA circuits,'' \emph{IOP SciNotes}, vol. 1, no. 2,
  pp. 025208, Aug. 2020, DOI.
  \href{https://doi.org/10.1088/2633-1357/abb0d7}{10.1088/2633-1357/abb0d7}.
\item \label{roffe_errorcorrect}
  J. Roffe, ``Quantum error correction: an introductory guide,''
  \emph{Contemporary Physics}, vol. 60, no. 3, pp. 226--245, Oct. 2019,
  DOI.
  \href{https://doi.org/10.1080/00107514.2019.1667078}{{10.1080/00107514.2019.1667078}}.
\item \label{shor_dechoerence}
  P. W. Shor, ``Scheme for reducing decoherence in quantum computer
  memory,'' \emph{Phys. Rev. A}, vol. 52, no. 4, pp. R2493, Oct. 1995,
  DOI.
  \href{https://doi.org/10.1103/PhysRevA.52.R2493}{10.1103/PhysRevA.52.R2493}.
\item \label{gottesman_stabilizer}
  D. Gottesman, ``Stabilizer codes and quantum error correction,'' Ph.D.
  dissertation, Dept. Physics, California Institute of Technology,
  Pasadena, CA, USA, 1997.
\item \label{kitaev_anyon}
  A. Y. Kitaev, ``Fault-tolerant quantum computation by anyons,''
  \emph{Annals of Physics}, vol. 303, no. 1, pp. 2--30, Jan. 2003, DOI.
  \href{https://doi.org/10.1016/S0003-4916(02)00018-0}{{10.1016/S0003-4916(02)00018-0}}.
\item \label{dennis_topomemory}
  E. Dennis, A. Kitaev, A. Landahl, and J. Preskill, ``Topological
  quantum memory,'' \emph{Journal of Mathematical Physics}, vol. 43, no.
  9, pp. 4452--4505, Aug, 2002, DOI.
  \href{https://doi.org/10.1063/1.1499754}{{10.1063/1.1499754}}.
\item \label{cai_bosonic}
  W. Cai, Y. Ma, W. Wang, C. L. Zou, and L. Sun, ``Bosonic quantum error
  correction codes in superconducting quantum circuits,''
  \emph{Fundamental Research}, vol. 1, no. 1, pp. 50--67, Jan, 2021,
  DOI.
  \href{https://doi.org/10.1016/j.fmre.2020.12.006}{{10.1016/j.fmre.2020.12.006}}.
\item \label{muller_VQE}
  J. Müller, W. Lavrijsen, C. Iancu, and W. de Jong, ``Accelerating
  noisy VQE optimization with Gaussian processes,'' in \emph{2022 IEEE
  International Conference on Quantum Computing and Engineering (QCE)},
  Broomfield, CO, USA, 2022, pp. 215--225, DOI.
  {10.1109/QCE53715.2022.00041}.
\item \label{gilliam_GAS} A. Gilliam, S. Woerner, and C. Gonciulea, “Grover adaptive search for constrained polynomial binary optimization,” Quantum, vol. 5, pp. 428, Apr. 2021, DOI. \href{https://doi.org/10.22331/q-2021-04-08-428}{10.22331/q-2021-04-08-428}.
\item \label{perdomo_protein}
  A. Perdomo-Ortiz, N. Dickson, M. Drew-Brook, G. Rose, and A.
  Aspuru-Guzik, ``Finding low-energy conformations of lattice protein
  models by quantum annealing,'' \emph{Scientific Reports}, vol. 2, no.
  1, pp. 1--7, Aug. 2012, DOI.
  \href{https://doi.org/10.1038/srep00571}{10.1038/srep00571}.
\item \label{setia_qubitreduce}
  K. Setia, R. Chen, J. E. Rice, A. Mezzacapo, M. Pistoia, and J. D.
  Whitfield, ``Reducing qubit requirements for quantum simulations using
  molecular point group symmetries,'' \emph{Journal of Chemical Theory
  and Computation}, vol. 16, no. 10, pp. 6091--6097, Aug. 2020, DOI.
  \href{https://doi.org/10.1021/acs.jctc.0c00113}{{10.1021/acs.jctc.0c00113}}.
\item \label{tran_mixedint}
  A. Tran, M. Tran, and Y. Wang. "Constrained mixed-integer Gaussian
  mixture Bayesian optimization and its applications in designing
  fractal and auxetic metamaterials," \emph{Structural and
  Multidisciplinary Optimization}, vol. 59, pp. 2131--2154, Jan. 2019,
  DOI.
  \href{https://doi.org/10.1007/s00158-018-2182-1}{10.1007/s00158-018-2182-1}.
\item \label{datta_discrete}
  D. Datta and J. R. Figueira, ``A real-integer-discrete-coded particle
  swarm optimization for design problems,'' \emph{Applied Soft
  Computing}, vol. 11, no. 4, pp. 3625--3633, Jun. 2011, DOI.
  \href{https://doi.org/10.1016/j.asoc.2011.01.034}{{10.1016/j.asoc.2011.01.034}}.
\item \label{cagnina_eng}
  L. C. Cagnina, S. C. Esquivel, and C. A. C. Coello, ``Solving
  engineering optimization problems with the simple constrained particle
  swarm optimizer,'' \emph{Informatica}, vol. 32, no. 3, pp. 319--326,
  Jan. 2008.
\end{enumerate}

\end{document}